\documentclass[iop]{emulateapj}
\usepackage{epstopdf}
\usepackage{epsfig}

\def\msun{\mbox{$M_\odot$}}
\def\lsun{\mbox{$L_\odot$}}
\def\rsun{\mbox{$R_\odot$}}

\def\teff{\mbox{$T_{\rm eff}$}}
\def\logg{\mbox{$\log g$}}

\def\ms{\mbox{m s$^{-1}$}}

\def\msy{\mbox{m s$^{-1}$ yr$^{-1}$}}
\def\ks{\mbox{km s$^{-1}$}}

\def\mjup{$M_{\rm Jup}$}

\def\vsini{$v \sin i$}
\def\msini{$M_{P} \sin{i}$}

\def\caii{\ion{Ca}{2}}
\def\shk{\mbox{$S_{\rm HK}$}}
\def\chinu{$\chi_{\nu}^{2}$~}
\def\plm{$\pm$\ }

\def\rms{RMS}

\def\innerp{$75.29$}

\def\tc{$t_{c}$}

\received{}
\accepted{2011/09/09}

\shorttitle{Planets Around HD~163607 \& HD~164509}
\shortauthors{Giguere et al.}
\begin{document}

\title{A High Eccentricity Component in the Double Planet System Around HD~163607 and a Planet Around HD~164509\altaffilmark{1}}

\author{Matthew J. Giguere\altaffilmark{2}, 
Debra A. Fischer\altaffilmark{2}, 
Andrew W. Howard\altaffilmark{3,4}, 
John A. Johnson\altaffilmark{5}, 
Gregory W. Henry\altaffilmark{6}, 
Jason T. Wright\altaffilmark{7,8}, 
Geoffrey W. Marcy\altaffilmark{3}, 
Howard T. Isaacson\altaffilmark{3},
Fengji Hou\altaffilmark{9}, 
Julien Spronck\altaffilmark{2}}

\begin{abstract}
We report the detection of three new exoplanets from Keck Observatory. HD~163607 is a metal-rich G5IV star with two planets. The inner planet has an observed orbital period of 75.29 $\pm$ 0.02 days, a semi-amplitude of 51.1 $\pm$ 1.4 \ms, an eccentricity of 0.73 $\pm$ 0.02 and a derived minimum mass of \msini = 0.77 $\pm$ 0.02 \mjup. This is the largest eccentricity of any known planet in a multi-planet system. The argument of periastron passage is 78.7 $\pm$ 2.0$^{\circ}$; consequently, the planet's closest approach to its parent star is very near the line of sight, leading to a relatively high transit probability of 8\%. The outer planet has an orbital period of 3.60 $\pm$ 0.02 years, an orbital eccentricity of 0.12 $\pm$ 0.06 and a semi-amplitude of 40.4 $\pm$ 1.3 \ms. The minimum mass is \msini = 2.29 $\pm$ 0.16 \mjup. HD~164509 is a metal-rich G5V star with a planet in an orbital period of 282.4 $\pm$ 3.8 days and an eccentricity of 0.26 $\pm$ 0.14. The semi-amplitude of 14.2 $\pm$ 2.7 \ms\ implies a minimum mass of 0.48 $\pm$ 0.09 \mjup. The radial velocities of HD~164509 also exhibit a residual linear trend of -5.1 $\pm$ 0.7 \ms\ per year, indicating the presence of an additional longer period companion in the system. Photometric observations demonstrate that HD~163607 and HD~164509 are constant in brightness to sub-millimag levels on their radial velocity periods.  This provides strong support for planetary reflex motion as the cause of the radial velocity variations.

\end{abstract}


\altaffiltext{1}{Based on observations obtained at the W. M. Keck Observatory, 
which is operated by the University of California and the California Institute of 
Technology. Keck time has been granted by NOAO and NASA.}

\altaffiltext{2}{Department of Astronomy, Yale University, 260 Whitney Ave, New Haven, CT 06511, USA}

\altaffiltext{3}{Department of Astronomy, University of California, Berkeley, Berkeley, CA 94720, USA}

\altaffiltext{4}{Space Sciences Laboratory, University of California, Berkeley, Berkeley, CA 94720, USA}

\altaffiltext{5}{Department of Astrophysics, California Institute of Technology, MC 249-17, Pasadena, CA 91125, USA}

\altaffiltext{6}{Center of Excellence in Information Systems, Tennessee State University,
3500 John A. Merritt Blvd., Box 9501, Nashville, TN 37209, USA}

\altaffiltext{7}{Department of Astronomy, 525 Davey Lab, The Pennsylvania State
ÊUniversity, University Park, PA 16802}

\altaffiltext{8}{Center for Exoplanets and Habitable Worlds, The
Pennsylvania State University, University Park, PA 16802, USA}

\altaffiltext{9}{Center for Cosmology and Particle Physics, Department of Physics, New York University, 4 Washington Place, New York, NY 10003, USA}
\ \

\section{Introduction}

Over the past 15 years more than 500 extrasolar planets have been detected in radial velocity, transit, microlensing and most recently, imaging surveys (http://exoplanet.eu, http://exoplanets.org).  In step with these discoveries, planet formation theories have made use of the observational constraints imposed by the ensemble of exoplanets \citep{2008ApJ...686..621F}. Correlations have been uncovered between parameters such as the chemical composition and mass of host stars and the probability of planets \citep{fv05, 2007ApJ...670..833J, 2010PASP..122..905J}. 

Based on the solar nebula model, there was an early expectation that planets would reside in nearly circular orbits, since eccentric orbits would have been quickly dampened through their interaction with dust in protoplanetary disks \citep{1993ARA&A..31..129L}. Therefore, the detection of significant eccentricity in exoplanet orbits and the unexpected parallel between the eccentricity distribution of planets and stars continues to be of strong interest \citep{1996ApJ...464L.147M, 1997ApJ...483..457C, 2008ApJ...686..621F}

One possibility is that non-coplanar orbits can be produced by planet-planet scattering in highly eccentric orbits \citep{2008ApJ...686..580C, 2008ApJ...686..603J}. Among transiting exoplanets where the Rossiter-McLaughlin (R-M) effect has been measured, approximately one third seem to be misaligned \citep{Triaud2010, 2010ApJ...718L.145W, 2010MNRAS.405.1867S}. Most of these transiting systems have low eccentricity orbits; only four known transiting systems have eccentricities greater than 0.4: HD~17156 \citep{2007ApJ...669.1336F, 2007A&A...476L..13B}, HD~80606 \citep{2001A&A...375L..27N, 2009MNRAS.396L..16F, 2009A&A...498L...5M}, HAT-P-2 b, \citep{2007ApJ...670..826B}, and CoRoT-10b \citep{2010A&A...520A..65B}. Using \textit{Kepler} data, \citet{Lissauer2011} find that nearly coplanar multi-planet systems of low mass planets in short-period orbits are quite common. However, based on recent observations of the few misaligned hot-Jupiter systems recently detected, \citet{2010ApJ...718L.145W} note that misaligned systems are preferentially detected around hot stars, but the underlying mechanism for misalignment is not clear. 

In this paper we present the detection of two new planetary systems orbiting metal rich stars. The first is a double-planet system orbiting HD~163607. The inner planet, with an orbital period of 75 days and an eccentricity of e $\sim$ 0.73, is the most eccentric planet detected in a multiplanet system to date. This eccentricity, combined with the argument of periastron passage of 78$^{\circ}$, makes the probability of transit 8\%, 3.7 times higher than it would be if it were in a circular orbit with the same orbital period. The orbital parameters are precise enough to provide good transit ephemeris predictions, and if this system transits, then a measurement of the R-M effect could shed light on orbital evolution for this system. The second system is a 0.5 \mjup\ planet orbiting HD~164509, with an orbital period of 245 days. 

In section \ref{sec:obs} we describe the observations and analysis of the two exoplanet systems, followed by a description of the photometric followup in section \ref{sec:pho}. In section \ref{sec:hd163}, we describe the properties, observations, orbital analysis, transit parameters and photometry concerning the host star and two substellar companions detected orbiting HD~163607. In section \ref{sec:hd164} we describe the detection of a jovian mass planet orbiting HD~164509. In section \ref{sec:disc} we summarize this paper and discuss the impact of these results. 

\section{Spectroscopic Observations and Reductions}
\label{sec:obs}

Doppler measurements for both HD~163607 and HD~164509 were obtained at the Keck Observatory using the HIRES spectrograph. The intial observations were made as part of the N2K program \citep{2005ApJ...620..481F}, which targeted high metallicity stars to detect short period jovian mass planets. Continued observations of these stars are now revealing longer period and multi-planet systems. 

Radial velocity measurements for the stars on our programs are determined with a forward-modeling process \citep{Marcy1992}. The ingredients in our model include an intrinsic stellar spectrum (ISS) of the star, a high resolution spectrum of the iodine cell and a model for the HIRES instrumental profile (IP). The ISS and the iodine spectrum are multiplied together, with the shift of the iodine spectrum as a free parameter to derive the radial velocity of the star. This product is then convolved with the IP model and binned on the HIRES pixel scale. Our Doppler model is broken up into approximately seven hundred 2~\AA\ chunks from 500 - 600 nm to accommodate variations in the IP across the detector. The single-measurement uncertainty is the weighted uncertainty in the mean radial velocities from these chunks.

%TABLE 1: STELLAR PARAMETERS:
\begin{deluxetable}{l l l l }
\tablenum{1}
\tablecaption{Stellar Parameters \label{tab:stellar}}
\tablewidth{0pt}
\tablehead{ & \colhead{HD~163607}     & \colhead{HD~164509} }
\startdata
Spectral type            & G5 IV                        & G5 V                     \\
$V$                            & 8.15                              & 8.24                \\
$M_V$                      & 3.96                          & 4.64   \\
$B-V$                        & 0.78                             &  0.66   \\
BC                              & -0.127                  &  -0.06         \\
Distance (pc)            & 69 (3)                   & 52 (3)             \\
$T_{\rm eff}$ (K)       & 5543 (44)        &  5922 (44) \\
\logg                           & 4.04 (6)                  &  4.44 (6)   \\
${\rm [Fe/H]}$           & 0.21 (3)                  &  0.21 (3)    \\
\vsini (\ks)                  & 1.49 (50)                 &  2.4 (5)        \\
$M_{\star}$ (\msun) & 1.09 (2)                & 1.13 (2) \\
$R_{\star}$ (\rsun)   & 1.63 (7)                    &  1.06 (3) \\
$L_{\star}$ (\lsun)    & 2.3 (2)                    &  1.15 (13) \\
Age (Gyr)                   &  8.6  (6)                    &   1.1 (1.0) \\
log R'$_{HK}$           & -5.01                    &   -4.88       \\
S$_{HK}$                   & 0.164                   &   0.18
\enddata                         
\end{deluxetable}

We derive stellar parameters (\teff, [Fe/H], \logg ~and \vsini) using the LTE spectral synthesis analysis software \textit{Spectroscopy Made Easy} (SME) \citep{vp96, vf05}. After generating an initial synthetic model, we iterate between the Y$^{2}$ isochrones \citep{d04} and SME model as described by \citet{2009ApJ...702..989V} until agreement in the surface gravity converges to 0.001 dex.  The stellar mass, luminosity and ages are derived from the Y$^{2}$ isochrones \citep{d04} and bolometric luminosity corrections are from \citet{Van03}.

Velocity jitter is a combination of astrophysical noise and systematic errors that can be misinterpreted as dynamical velocities. Starspots are one source of stellar jitter. As stars rotate, their spots will rise and set over the approaching and receding stellar limbs shifting the centroids of spectral lines and leading to confusion with dynamical Doppler line shifts. \citet{if10} measured emission in the \caii\ line cores and \shk\ values to derive $\log{R'_{HK}}$, the ratio of emission in the core of the \caii\ lines to the photospheric values. They have derived astrophysical jitter measurements as a function of \bv\ color, luminosity class, and excess \shk\ values and we adopt those stellar jitter measurements for the stars in this paper. 

To carry out Keplerian modeling of RV and astrometric data, we developed a software package by the name of \textit{Keplerian Fitting Made Easy (KFME)}. \textit{KFME} was programmed in IDL and the graphical user interface was inspired by the Systemic Console \citep{2010ApJ...718..543M}. The GUI was initially developed to fit synthetic radial velocity and astrometry data for the Space Interferometry Mission (SIM) \citep{2010EAS....42..191T}. While SIM has been officially discontinued, there is still the possibility of using the full potential of \textit{KFME}, given the recent discovery of an exoplanet via astrometry \citep{Muterspaugh2010}. However, \textit{KFME} has been used here to fit radial velocity data alone. 

\textit{KFME} displays the data set and allows the user to adjust initial parameters for up to seven planets. Functions are included for periodogram analysis, Levenberg-Marquardt fitting, analysis of false alarm probability, and Bootstrap Monte Carlo analysis to determine parameter uncertainty. \textit{KFME} also offers an automated option that cycles through several values of $\omega$ and $T_{P}$, retaining the lowest \chinu\ as the best fit solution. 

The false alarm probability (FAP) in \textit{KFME} is calculated with a Monte Carlo simulation. Before the MC synthetic data sets are created to calculate the FAP for any single planet, any linear trend that is present and/or the best-fit Keplerian model for additional planets are subtracted from the velocities. Then, for each MC trial, the observation times are kept fixed, while the associated velocities are scrambled (replacement, with redraw of the same value is allowed). Each MC trial of scrambled velocities is then blindly fit with a Levenberg-Marquardt least-squares minimization algorithm with a Keplerian model. The lowest \chinu\ for each test case is stored in an array. After N trials (where N is set by the desired precision in the FAP) the array of \chinu\ values is sorted and a comparison with the original \chinu\ of the unscrambled velocities is made to this ranked set. The FAP assesses how frequently the \chinu\ of the scrambled velocities is lower than the \chinu\ of the original, unscrambled velocities. For example, if the scrambled data yielded periodograms with peaks of comparable height as the original peak in 10 out of 100 trials, the FAP of 10/100 = 0.1 reflects the fact that spurious signals could have occurred 10\% of the time. 

For multiplanet systems, signals can be individually displayed. \textit{KFME} allows fitting for systematic velocity offsets (e.g., different observatories, different detectors, etc.) and the inclusion of jitter, added in quadrature to the formal errors. After fitting for one or more planets, residual velocities can be displayed and analyzed. 

One advangage of \textit{KFME} is that it engages the human brain, which is often good at discerning global patterns, to approximate the initial conditions for generating a Keplerian model with a Levenberg-Marquardt fitting algorithm. The Levenberg-Marquardt algorithm then polishes this approximate solution into a low \chinu\ fit. The best fit can then be used to generate transit parameters (ingress, egress, duration, time of center transit, and probability). \textit{KFME} also collects all of our orbital modeling tools into one package. \textit{KFME} can be downloaded at exoplanets.astro.yale.edu/KFME, and requires IDL in order to run. 

\section{Photometric Observations and Reductions}
\label{sec:pho}

We acquired photometric observations of HD~163607 and HD~164509 
with the T12 0.80~m automatic photometric telescope (APT) at Fairborn 
Observatory.  The T12 APT and its two-channel photometer measure photon 
count rates simultaneously through Str\"omgren $b$ and $y$ filters.  
T12 is essentially identical to the T8 0.80~m APT described in 
\citet{henry1999}. 

The two program ($P_{g}$) stars HD~163607 and HD~164509 were each observed 
differentially with respect to two nearby comparison stars ($C1$ and $C2$).  
The two comparison stars for HD~163607 were HD~169352 ($V=8.01$, $B-V=0.44$, F2) 
and HD~165700 ($V=7.79$, $B-V=0.46$, F8); comparison stars for HD~164509 were 
HD~166073 ($V=6.99$, $B-V=0.45$, F7~IV) and HD~165146 ($V=7.57$, $B-V=0.45$, 
F0).  The differential magnitudes $P_{g}-C1$, $P_{g}-C2$, and $C2-C1$ were computed 
from each set of differential measures.  The observations were corrected 
for extinction and transformed to the Str\"omgren photometric system.  
To improve the precision of our brightness measurements, we averaged the 
$b$ and $y$ differential magnitudes into a single $(b+y)/2$ ``passband'', 
which we designate $by$.  Typical precision of a single $by$ observation 
is $~\sim0.0015-0.0020$ mag, as measured for pairs of constant stars.  
\citet{henry1999} provides additional details on the operation of the APT, 
observing and data reduction procedures, and precision of the data.

\citet{qhs+2001} and \citet{psch2004} have demonstrated how rotational 
modulation of starspots on active stars can result in periodic radial 
velocity variations that mimic the presence of a planetary companion.  
Thus, the precise APT brightness measurements are valuable for 
distinguishing between activity-related RV changes and true reflex motion 
of a star caused by a planetary companion.  

\section{HD~163607}
\label{sec:hd163}
\label{sec:stellar}

%TABLE 2: RADIAL VELOCITY 163607:
\begin{deluxetable}{rrrr}
\tablenum{2}
\tablecaption{Radial Velocities for HD~163607 \label{tab:rvs}}
\tablewidth{0pt}
\tablehead{\colhead{JD}  & \colhead{RV}  
& \colhead{$\sigma_{RV}$} & \colhead{$S_{HK}$}   \\
\colhead{-2440000}  & \colhead{(\ms)}   & \colhead{(\ms)}  } 
\startdata
13570.8646  &  -23.74  &  1.18 & 0.164  \\  
13575.9339  &  -13.41  &  1.24 & 0.161  \\  
13576.8468  &  -7.42  &  1.30 & ------  \\  
14247.9039  &  38.39  &  1.13 & 0.164  \\  
14249.8472  &  46.49  &  1.27 & 0.164  \\  
14250.9825  &  47.65  &  1.18 & 0.165  \\  
14251.9291  &  52.15  &  1.13 & 0.164  \\  
14285.8917  &  -7.61  &  0.97 & 0.164  \\  
14318.9150  &  23.03  &  0.83 & 0.164  \\  
14339.8773  &  -44.90  &  1.05 & 0.164  \\  
14343.8563  &  -41.10  &  0.93 & 0.165  \\  
14345.7946  &  -31.49  &  1.37 & 0.167  \\  
14345.8208  &  -33.73  &  1.05 & 0.164  \\  
14549.0883  &  -6.88  &  1.18 & 0.164  \\  
14634.9349  &  14.16  &  1.00 & 0.164  \\  
14639.0081  &  -84.13  &  1.00 & 0.168  \\  
14641.9689  &  -90.54  &  1.12 & 0.164  \\  
14674.8896  &  -49.94  &  1.02 & 0.165  \\  
14689.9213  &  -36.62  &  1.12 & 0.165  \\  
15016.0415  &  -79.07  &  0.96 & 0.164  \\  
15041.9308  &  -39.20  &  1.14 & 0.165  \\  
15044.0017  &  -42.11  &  1.08 & 0.164  \\  
15073.7742  &  -0.64  &  1.11 & 0.165  \\  
15082.7333  &  26.03  &  1.09 & 0.163  \\  
15111.7501  &  -35.94  &  1.20 & 0.165  \\  
15135.7005  &  -2.87  &  1.13 & 0.164  \\  
15163.6885  &  17.28  &  1.30 & 0.161  \\  
15172.6923  &  -42.17  &  1.19 & 0.162  \\  
15229.1489  &  46.57  &  1.06 & 0.163  \\  
15232.1671  &  56.60  &  1.14 & 0.161  \\  
15256.1669  &  -7.61  &  1.16 & 0.162  \\  
15261.0850  &  -1.71  &  1.13 & 0.164  \\  
15285.0537  &  28.62  &  1.13 & 0.154  \\  
15313.9685  &  50.43  &  1.13 & 0.163  \\  
15319.0701  &  -17.97  &  1.22 & 0.161  \\  
15351.1269  &  11.69  &  1.02 & 0.163  \\  
15373.8050  &  43.04  &  1.09 & 0.164  \\  
15399.9669  &  -12.62  &  1.02 & 0.163  \\  
15436.7391  &  27.49  &  1.07 & 0.157  \\  
15455.7818  &  53.65  &  1.00 & 0.164  \\  
15486.7739  &  1.75  &  1.04 & 0.164  \\  
15500.6979  &  10.70  &  1.16 & 0.164  \\  
15521.7059  &  32.83  &  1.04 & 0.163  \\  
15606.1749  &  40.56  &  1.09 & 0.164  \\  
15606.1796  &  41.72  &  1.11 & 0.163  \\  
15613.1249  &  61.89  &  1.84 & 0.158  \\  
15634.0875  &  -17.56  &  1.03 & 0.164  \\  
15637.0856  &  -17.19  &  1.06 & 0.163  \\  
15669.0353  &  14.99  &  1.07 & 0.164  \\  
15671.0957  &  18.70  &  1.08 & 0.164  \\  
15700.8908  &  -44.97  &  1.15 & 0.164  \\  
15728.9113  &  -16.27  &  1.16 & 0.164  \\  
15763.7659  &  40.29  &  1.00 & 0.164  \\  
\enddata
\end{deluxetable}

HD~163607 (HIP~87601) is a G5 subgiant at a distance of 69 $\pm$ 3 pc calculated from the \textit{Hipparcos} parallax measurement \citep{esa97} and revised catalog \citep{van08}. We adopted the \textit{Hipparcos} V-band magnitude and color of V = 8.15 and \bv\ = 0.78. With a bolometric correction of -0.127, this gives the absolute visual magnitude of $M_{V} = 3.96$. 

An iodine-free ``template" spectrum of HD~163607 was analyzed by iterating SME models with Y$^{2}$ isochrones to derive the following stellar parameters:  \teff = 5543 $\pm$ 44 K,  [Fe/H] = 0.21 $\pm$ 0.03 dex, projected stellar rotational velocity, \vsini = 1.5 $\pm$ 0.5 \ks, \logg = 4.04 $\pm$ 0.06. The isochrone analysis also yielded an age of 8.6 $\pm$ 0.6 Gyr, a stellar radius 1.63 $\pm$ 0.07 R$_{\sun}$ and a luminosity of 2.3 $\pm$ 0.2 L$_{\sun}$. HD~163607 has low chromospheric activity with $\log{R'_{HK}} = -5.01$ and an estimated stellar jitter of 2.6 \ms. We do not see a correlation between activity and the measured radial velocities. The stellar properties for HD~163607 are summarized in Table \ref{tab:stellar}.

\begin{figure}[h!]
\epsfig{file=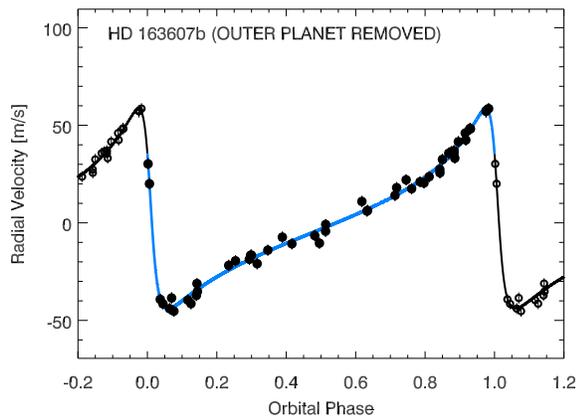,width=1.1\linewidth,clip=}
\caption{Phase-folded radial velocities for HD~163607b, the inner planetary companion in this system, which has an orbital period of 75.29 days. The theoretical velocities from the outer planet have been subtracted from the observations and the curve shown is the best-fit model for this system.}
\label{figHD163607rvfitb}
\end{figure}

\subsection{Orbital Solution}
\label{sec:163orb}

Observations of HD~163607 began in July of 2005. The 51 observations of this star are listed in Table \ref{tab:rvs} and have a median velocity precision of 1.11 \ms.  We analyzed the radial velocity data for HD~163607 using \textit{KFME} and included jitter of 2.6 \ms. The single planet model had a period of 1500 days, however, the residuals to this fit had an rms of 31 \ms\ and significant power at 75 days. The addition of a second planet in the Keplerian model reduced the \chinu\ fit to 1.03 with an rms of 2.9 \ms. Once the best-fit two planet Keplerian model was attained using \textit{KFME}, a Bootstrap Monte Carlo routine that is also built into \textit{KFME} was used to derive uncertainties for the orbital parameters. The Bootstrap Monte Carlo routine, similar to the routine employed to calculate the FAP described in section \ref{sec:obs}, subtracts the best-fit model from the velocities, scrambles the residuals and their associated uncertainties, adds them back to the model and refits. The uncertainties quoted in this section and listed in Table \ref{tab:orb} come from 10$^{3}$ realizations of this Bootstrap Monte Carlo routine. 

\begin{figure}[h]
\epsfig{file=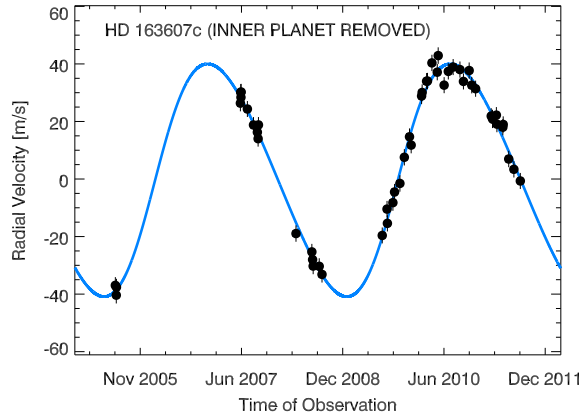,width=1.1\linewidth,clip=}
\caption{Radial velocities for HD~163607 with the theoretical velocities for the inner planet removed from the observations. The curve shown is the best-fit Keplerian model for HD~163607c. }
\label{figHD163607rvfitc}
\end{figure}

The inner planet has a best-fit period, $P$, of \innerp ~$\pm$ 0.02 days and velocity amplitude, $K$, of $51.1 \pm 1.4$ \ms. The best-fit eccentricity, $e$, for this planet is $0.73 \pm 0.02$. Assuming a stellar mass of  1.09 \msun, the derived minimum planet mass is \msini = 0.77 $\pm$ 0.04 \mjup. The high eccentricity and favorable argument of periastron passage $\omega$, conspire to increase the probability of transit to 8\%; much larger than it would have been if the inner planet was in a circular orbit. The observed radial velocities for the inner planet as a function of orbital phase are shown in Figure \ref{figHD163607rvfitb}. In this figure the Keplerian model for the outer planet has been removed and the Keplerian model for the inner planet is plotted as the solid line. 

\begin{figure}
\epsfig{file=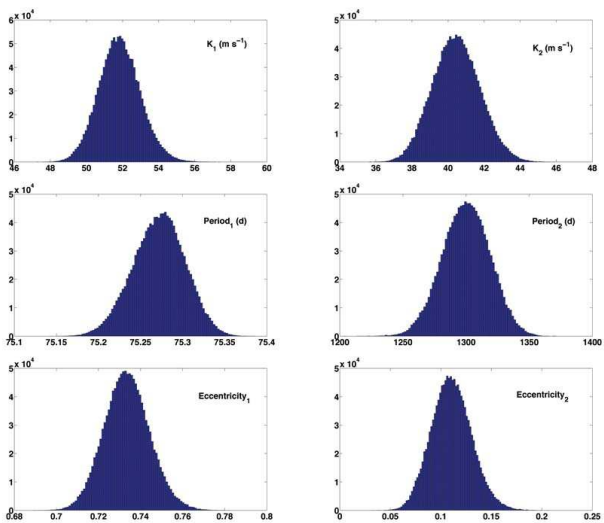,width=1.1\linewidth,clip=}
\caption{MCMC histograms showing the posterior PDFs for several orbital parameters for HD~163607b (left) and HD~163607c (right). The very narrow PDFs for HD~163607b reflect the excellent phase coverage near periastron passage as seen in Figure \ref{figHD163607rvfitb}.}
\label{fig:bayes163}
\end{figure}

\begin{figure*}
\begin{center}
\epsfig{file=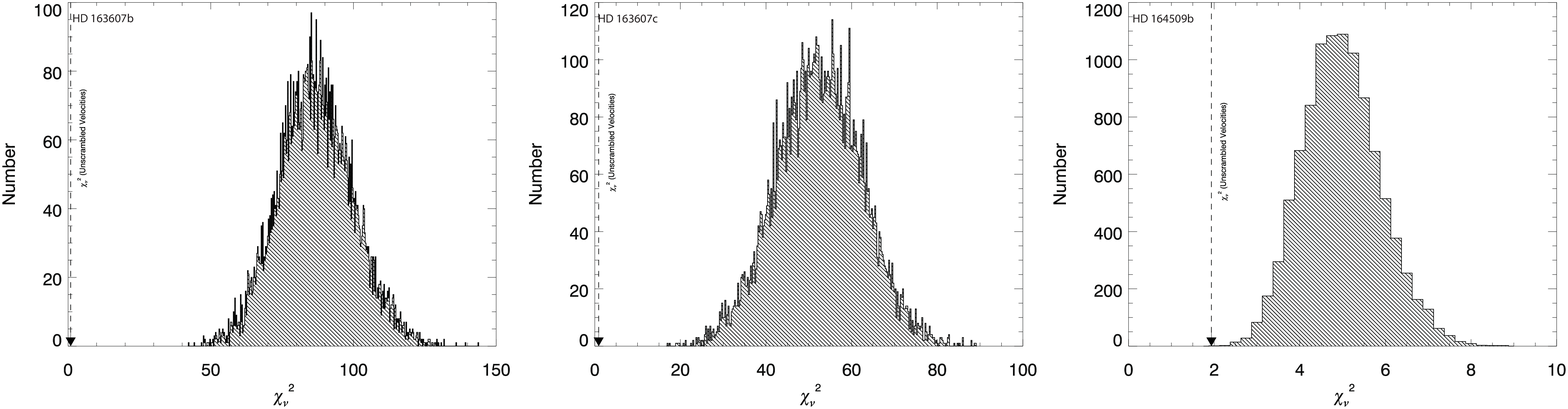, width=7in}
%\epsscale{1}
\caption{To determine the false alarm probability (FAP) for each planet, the velocities were scrambled 10,000 times and each set of velocities was blindly fit with a Keplerian model. The above figures show the distributions of these realizations as a function of $\chi_{\nu}^{2}$ for HD~163607b (left), HD~163607c (middle) and HD~164509b (right). The dashed arrow shows the $\chi_{\nu}^{2}$ for the unscrambled velocities, which is well to the left of each distribution. The FAP for each planet was $\le 0.01\%$.}
\label{fig:FAPs}
\end{center}
\end{figure*}

The outer planet, HD~163607c, has a best-fit period,  $P$, of 1314 $\pm$ 8 days, a velocity semi-amplitude of 40.4 $\pm$ 1.3 \ms\, and an eccentricity, $e$, of 0.12 $\pm$ 0.06. The derived mass is 2.29 $\pm$ 0.16 \mjup\ and the semi-major axis of the orbit is 2.42 $\pm$ 0.01 AU. In Figure \ref{figHD163607rvfitc}, the Keplerian model fit to the data for HD~163607c is shown with the inner planet subtracted from the Doppler measurements. MCMC simulations to confirm our calculation of the best-fit orbital parameters were carried out using the methods of \citet{Hou2011}, and the resulting posterior PDFs for several orbital parameters can be seen in Figure \ref{fig:bayes163}. There was excellent agreement between the MCMC results and the frequentist approach using \textit{KFME}. The orbital parameters of both planets obtained using \textit{KFME} are summarized in Table \ref{tab:orb}. 

%TABLE 3: ORBITAL PARAMETERS:
%UPDATED: 20110725

\begin{deluxetable*}{lrrr}
\tablenum{3}
\tablecaption{Orbital parameters for the three planets described in this work. \label{tab:orb}}
\tablewidth{3in}
\tablehead{
\colhead{Parameter}  & 
\colhead{HD 163607b} & 
\colhead{HD 163607c} &
\colhead{HD 164509b}
} 
\startdata
P(d)                                                     & 75.29 $\pm$ 0.02                 & 1314 $\pm$ 8          & 282.4 $\pm$ 3.8 \\
T$_{P}$(HJD\tablenotemark{*})       & 14185.00 $\pm$ 0.24          & 15085 $\pm$ 880     & 15703 $\pm$ 30 \\
t$_{C}$(HJD\tablenotemark{*})        &  15841.59 $\pm$ 0.24     &      17074 $\pm$ 15      & 15498 $\pm$ 22 \\
e                                                           &  0.73 $\pm$ 0.02                  &  0.12 $\pm$ 0.06    & 0.26 $\pm$ 0.14 \\
$\omega$                                           &  78.7 $\pm$ 2.0                     & 265 $\pm$ 93         & 324 $\pm$ 110 \\
K  (m s$^{-1}$)                                   &  51.1 $\pm$ 1.4                    & 40.4 $\pm$ 1.3        & 14.2 $\pm$ 2.7 \\
a (AU)                                                  &  0.36 $\pm$ 0.01                  & 2.42 $\pm$ 0.01     & 0.875 $\pm$ 0.008 \\
M $\sin{i}$  (\mjup)                            &  0.77 $\pm$    0.04               & 2.29 $\pm$ 0.16     & 0.48  $\pm$ 0.09 \\
\hline
$\gamma$ (m s$^{-1}$)                    & -15.7 $\pm$ 0.5     &                                         &   8.9 $\pm$ 2.1\\
dvdt (m s$^{-1}$ yr$^{-1}$)              & 0                                &                                        &  -5.1 $\pm$ 0.7 \\
N$_{\textrm{obs}}$                            &   51                                 &                                  & 41 \\
Jitter (m s$^{-1}$)                              &  2.6                                  &                                  & 3.2 \\
rms  (m s$^{-1}$)                               &   2.9                                     &                              & 4.9 \\
$\chi_{\nu}^{2}$                                 &   1.03                                        &                         & 2.04 \\
\enddata                        
\tablenotetext{*}{HJD - 2,440,000}
\end{deluxetable*}

Similar to the \textit{scrambled velocities} method described in \citet{marcy2005}, the false alarm probabilities (FAPs) for each planet were estimated by creating 10$^{4}$ synthetic data sets by drawing, with replacement, velocities and their associated errors from the data, and placing these at the actual observation times. A Levenberg-Marquardt least-squares minimization to a Keplerian model was then performed for each synthetic data set, and the \chinu\ distributions for both planets orbiting HD~163607 are shown in Figure \ref{fig:FAPs}. The \chinu\ for the unscrambled data set is indicated by the downward pointing arrow in each plot, which is much lower than the \chinu\ of any of the 10$^{4}$ scrambled velocities for both planets in this system. This results in a FAP of $\le 0.01\%$, indicating that the observed signal is not due to noise, but rather a coherent periodicity that is well-fit with a Keplerian model.

%TABLE 4: PHOTOMETRIC PARAMETERS:
\begin{center}
%\begin{landscape}
\begin{deluxetable*}{ccccccccc}
%\rotate

\tabletypesize{\tiny}
\tablewidth{7in}
\tablenum{4}
\tablecaption{SUMMARY OF PHOTOMETRIC OBSERVATIONS FROM THE T12 APT \label{tab:pho}}
\tablehead{
\colhead{Program} & 
\colhead{Date Range} & \colhead{Duration} & \colhead{} & 
\colhead{$\sigma{(C2-C1)}_{by}$} & \colhead{$\sigma{(P_{g}-C1C2)}_{by}$} &
\colhead{} & \colhead{Orbital Period} & \colhead{Semiamplitude} \\
\colhead{Star} &
\colhead{(HJD $-$ 2,400,000)} & \colhead{(days)} & \colhead{$N_{obs}$} & 
\colhead{(mag)} & \colhead{(mag)} & \colhead{Planet} & 
\colhead{(days)} & \colhead{(mag)} \\
\colhead{(1)} & \colhead{(2)} & \colhead{(3)} & \colhead{(4)} & 
\colhead{(5)} & \colhead{(6)} & \colhead{(7)} & \colhead{(8)} & 
\colhead{(9)}
}
\startdata
 HD 163607 & 54381--55499 & 1118 & 156 & 0.0019 & 0.0015 & b &  75.29 & $0.0002\pm0.0002$ \\
           &              &      &     &        &        & c &   1314 & $0.0006\pm0.0003$ \\
 HD 164509 & 54377--55479 & 1102 & 168 & 0.0018 & 0.0018 & b & 282.4 & $0.0006\pm0.0002$ \\
\enddata

\end{deluxetable*}
\end{center}

%\end{landscape}

\subsection{Transit Prediction}
\label{sec:163transit}

\begin{figure}[h]
\epsfig{file=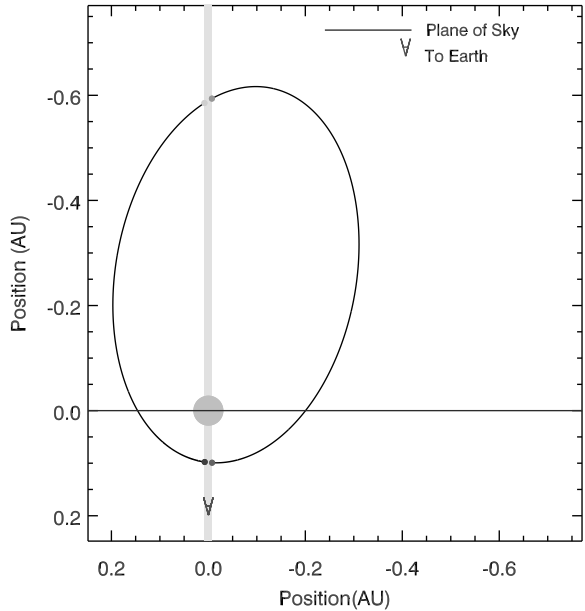,width=1.1\linewidth,clip=}
\caption{A view of the HD~163607b planetary orbit. The direction of the observer is indicated in the bottom of the illustration. The dots indicate the ingress and egress for the primary and secondary transits.}
\label{figHD163607pos}
\end{figure}

The proximity of hot Jupiters to their host stars results in an increased transit probability compared to longer period planets. Although the orbital period of HD~163607b is 75 days, the high orbital eccentricity moves the planet close to the star during periastron passage. Since the periastron passage of HD 163607b is serendipitously oriented very close to our line of site, the probability of transit is much larger than it would be if the planet was in a circular orbit. To illustrate this point, the orbital configuration for HD~163607b is shown in Figure \ref{figHD163607pos}. The calculation of the time of center transit, ingress and egress are particularly helpful for planning photometric follow-up for longer period planets. Below we describe a technique for calculating parameters to aid in photometric followup.

\subsubsection{Transit Center, Probability and Duration}

The term \textit{time of center transit}, \tc, as defined here is the time when the planet is nearest the center of the disk of its parent star as seen from Earth. We calculated \tc\ by using the best-fit orbital parameters to produce an array of Cartesian coordinates, giving the position of the planet relative to the host star in the inclination-projected orbital plane. The inclination-projected orbital plane was then rotated into the plane defined by the Earth, ascending, and descending nodes using the Thiele-Innes coordinates. The ingress and egress were then derived using the coordinates where the planet entered and exited the disk of the parent star. The true anomaly corresponding to each position can then be found and in turn the mean anomaly calculated. The mean anomaly, $M$, can then be used to find the time since periastron passage using Equation \ref{eqn:mean}:

\begin{equation}
t = \frac{PM}{2 \pi}
\label{eqn:mean}
\end{equation}

\noindent where $P$ is the orbital period, $M$ is the mean anomaly and $t$ is the time since periastron passage. Finding the difference in the mean anomalies for egress and ingress gives the transit duration, $t_{d}$. The transit center is then given by equation \ref{eqn:tcen}:

\begin{equation}
t_{c} = t + T_{P}.
\label{eqn:tcen}
\end{equation}

\begin{figure}[h]
\epsfig{file=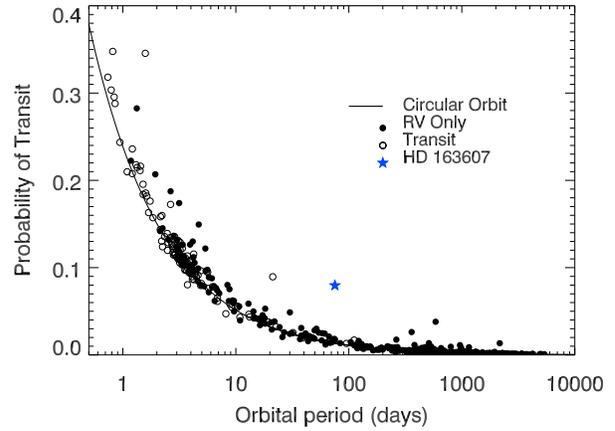,width=1.1\linewidth,clip=}
\caption{The probability of transit versus orbital period for the known distribution of exoplanets. The solid line is the probability for a circular orbit of a Jupiter radius planet around a one solar radius star. The filled circles represent planets detected through the radial velocity method that have not been observed to transit. The open circles represent planets that are known to transit and the star symbol represents HD~163607b.}
\label{fig:probper}
\end{figure}

Analytically, the time of central transit follows from the definition of the true anomaly and the argument of periastron, as described by \citet{2005ApJ...627.1011T, 2007MNRAS.380.1488K, 2009PASP..121.1386K}. The probability that the planetary companion will transit is given by \citet{2007prpl.conf..701C} and is reproduced in Equation \ref{eqn:prob}. The last factor in square brackets in this equation shows that HD~163607 is 3.7 times more likely to transit its parent star than if it were in a circular orbit with the same semi-major axis. 

\begin{equation}
P = 4.5*10^{-3}\left(\frac{1 \mathrm{AU}}{a} \right)\left(\frac{R_{\star}+R_{pl}}{R_{\sun}} \right)
\left[ \frac{1 + e\cos{\left(\frac{\pi}{2}-\omega \right)}}{1-e^{2}} \right]
\label{eqn:prob}
\end{equation}

In Equation \ref{eqn:prob}, $a$ is the semi-major axis, $R_{\star}$ is the radius of the star, $R_{pl}$ is the radius of the planet, $e$ is the eccentricity of the orbit and $\omega$ is the argument of periastron passage, which is the angle between the ascending node and the position of the planet at periastron. Figure \ref{fig:probper} shows the probability of transit for the known distribution of planets when taking the eccentricity and argument of periastron passage into account. The solid line shows the probability of transit as a function of orbital period for a Jupiter radius exoplanet transiting a solar radius star in a circular orbit. The filled circles are planets that have only been detected through the radial velocity method and the open circles are planets that have been observed to transit their parent stars. For both the filled and unfilled circles, the parent star's radius is set to solar and each planet's radius is set to the radius of Jupiter. The blue star shows where the inner companion orbiting HD~163607 falls on this plot, illustrating how the favorable orientation and high eccentricity enhance the probability of transit.

Using the equations described in this section we calculated the next time of center transit to be $t_{c}$ = 2455841.59 $\pm 0.24$ UT, the duration is roughly 5 hours, the depth is $\sim$~0.4\% ($\sim$4 mmag) and the probability of transit is 8\% for HD~163607b. The uncertainty in $t_{c}$ is determined while calculating the uncertainty in the orbital parameters. A bootstrap monte carlo routine is employed that subtracts the best fit Keplerian model, scrambles the residuals and their associated errors, adds these residuals back to the theoretical curve and refits. The new best fit orbital parameters are then used to determine $t_{c}$. The standard deviation of the resulting $t_{c}$ array after 10$^{3}$ realizations is what is stated for the uncertainty in $t_{c}$. The predicted transit window is defined in this work as the time between ingress and egress plus or minus one sigma. None of our radial velocity spectroscopic observations happened to fall within the predicted transit window. 

\subsection{Photometry}
Photometric results for HD~163607 are given in the first two rows of 
Table \ref{tab:pho}.  The observations were acquired between 2007 October 3 and 
2010 October 29.  Column 5 gives the standard deviation of the comparison star 
differential magnitudes $C2-C1$ as 0.0019 mag.  This is typical for 
constant stars with this telescope.  Periodogram analysis of the comparison 
star observations did not detect any significant periodicity between one 
and 200 days, so both comparison stars are constant to the level of the 
photometric precision.  To improve the precision of the HD~163607 
observations, we computed the $P_{g}-C1$ and $P_{g}-C2$ differential magnitudes 
in the $(b+y)/2$ passbands and then took the mean of those two 
differential magnitudes.  This results in differential magnitudes in the 
sense HD~163607 minus the mean brightness of the two comparison stars, which 
we designate as $P_{g}-C1C2$.  The standard deviation of the $P_{g}-C1C2$ 
observations is given in column 6 of Table \ref{tab:pho}.  The small value of 
0.0015 mag and the absence of any significant periodicity in the $P_{g}-C1C2$ 
observations indicates that HD~163607 is also constant to the limit of 
precision. 

We computed a least-squares sine fit of the $P_{g}-C1C2$ observations phased 
separately to the 75.29- and 1314-day periods of planets b and c, 
respectively.  Planet b exhibits a semiamplitude of $0.0002\pm0.0002$ mag, 
which is consistent with zero to high precision.  Planet c has a slightly 
larger semiamplitude of $0.0006 \pm 0.0003$ mag. We note that our 
observations do not quite cover one full cycle of the 1314-day RV 
variations.  However, given the very low standard deviation of the 
$P_{g}-C1C2$ observations, we conclude that HD~163607 is also constant to 
high precision on the period of planet c.  These very tight limits of 
brightness variability on the 75.29- and 1314-day RV periods strongly 
support the interpretation that the RV variations in HD~163607 are due to stellar reflex 
motion in gravitational response to planets b and c.  Our photometric
observations are too few for transit searches for the two planets.

%RADIAL VELOCITIES 164509:
\begin{deluxetable}{rrrr}
\tablenum{5}
\tablecaption{Radial Velocities for HD~164509 \label{tab:rvs164509}}
\tablewidth{0pt}
\tablehead{\colhead{JD}  & \colhead{RV}  
& \colhead{$\sigma_{RV}$} & \colhead{$S_{HK}$}   \\
\colhead{-2440000}  & \colhead{(\ms)}   & \colhead{(\ms)}  } 
\startdata
13570.8176  &  0.87  &  1.47 & 0.187  \\  
13576.0255  &  16.38  &  1.56 & 0.180  \\  
13576.8579  &  14.03  &  1.43 & ------  \\  
14251.0462  &  4.61  &  1.28 & 0.179  \\  
14318.8554  &  27.68  &  1.14 & 0.182  \\  
14339.7544  &  20.51  &  1.25 & 0.184  \\  
14343.8855  &  21.89  &  1.16 & 0.183  \\  
14633.9129  &  12.49  &  1.33 & 0.183  \\  
14634.8953  &  12.71  &  1.37 & 0.184  \\  
14635.9216  &  10.06  &  1.28 & 0.183  \\  
14637.0538  &  3.71  &  1.36 & 0.183  \\  
14637.9282  &  11.22  &  1.22 & 0.182  \\  
14638.9721  &  5.64  &  1.29 & 0.182  \\  
14639.9825  &  4.53  &  1.18 & 0.180  \\  
14641.9386  &  4.27  &  1.36 & 0.179  \\  
14644.0757  &  3.78  &  1.45 & 0.178  \\  
14674.8405  &  4.47  &  1.29 & 0.183  \\  
14688.8406  &  -7.39  &  1.36 & 0.182  \\  
14930.0681  &  7.13  &  1.38 & 0.192  \\  
14930.0700  &  5.49  &  2.16 & 0.188  \\  
14956.1303  &  -0.07  &  1.27 & 0.189  \\  
14964.0480  &  6.96  &  1.23 & 0.189  \\  
14985.1079  &  -1.27  &  1.47 & 0.190  \\  
15016.9568  &  -16.55  &  1.27 & 0.188  \\  
15019.0135  &  -18.16  &  1.50 & 0.186  \\  
15041.9070  &  -10.20  &  1.33 & 0.187  \\  
15043.7946  &  -11.64  &  1.37 & 0.187  \\  
15048.7836  &  -14.61  &  1.41 & 0.188  \\  
15077.7308  &  -11.39  &  1.18 & 0.182  \\  
15106.7446  &  2.18  &  1.29 & 0.183  \\  
15135.7359  &  8.32  &  1.33 & 0.185  \\  
15229.1718  &  -1.50  &  1.32 & 0.186  \\  
15232.1564  &  -9.79  &  1.29 & 0.180  \\  
15256.1429  &  -17.10  &  1.25 & 0.172  \\  
15286.1172  &  -14.85  &  1.29 & 0.185  \\  
15350.9342  &  -15.92  &  1.19 & 0.185  \\  
15380.7878  &  -13.97  &  1.37 & 0.179  \\  
15455.7291  &  3.08  &  1.21 & 0.182  \\  
15486.7349  &  -4.27  &  1.30 & 0.181  \\  
15500.7174  &  1.43  &  1.28 & 0.182  \\  
15521.6869  &  -12.83  &  1.46 & 0.178  \\  
15522.6823  &  -15.26  &  1.42 & 0.178  \\  
15607.1406  &  -24.51  &  1.25 & 0.183  \\  
15636.1381  &  -9.79  &  2.91 & 0.179  \\  
15668.0865  &  -17.46  &  1.14 & 0.178  \\  
15669.0153  &  -14.03  &  1.27 & 0.176  \\  
15671.0647  &  -7.97  &  1.34 & 0.177  \\  
15701.1294  &  3.91  &  1.37 & 0.178  \\  
15723.9194  &  3.25  &  1.42 & ------   \\  
\enddata
\end{deluxetable}

\section{HD~164509}
\label{sec:hd164}

HD~164509 (HIP~88268) is a G5 main sequence star at a distance of 52 $\pm$ 3 pc calculated from the \textit{Hipparcos} parallax measurement \citep{esa97, van08}. We adopted the V-band magnitude and color from the revised \textit{Hipparcos} catalog of V = 8.24 and B - V of 0.66, and derive an absolute visual magnitude of $M_{V} = 4.64$, which includes a bolometric correction of -0.06. Spectroscopic analysis of HD~164509 yields:  \teff = 5922 $\pm$ 44 K, [Fe/H] = 0.21 $\pm$ 0.03 dex,  \vsini = 2.4 $\pm$ 0.5 \ks\ and \logg = 4.44 $\pm$ 0.06. The isochrone analysis yields a mass of 1.13 $\pm$ 0.02 \msun, an age of 1.1 $\pm$ 1 Gyr, a stellar radius of 1.06 $\pm$ 0.03 R$_{\sun}$ and a luminosity of 1.15 $\pm$ 0.13 L$_{\sun}$.  The star has modest chromospheric activity, with $\log{R'_{HK}} = -4.88$ and \citet{if10} estimate a stellar jitter of 3.2 \ms. The stellar properties described above for HD~164509 are summarized in Table \ref{tab:stellar}.

\subsection{Orbital Solution}
\label{sec:orb}
Observations of HD~164509 began in July of 2005 at Keck Observatory with the HIRES spectrometer; and 41 observations of this star now span a time baseline of 5 years. The observations are listed in Table \ref{tab:rvs164509} and the median velocity error for HD~164509 is 1.32~\ms. We carried out Keplerian modeling for HD~164509 using \textit{KFME} after adding jitter of 3.2 \ms\ in quadrature with the formal velocity errors. 

\begin{figure}
\epsfig{file=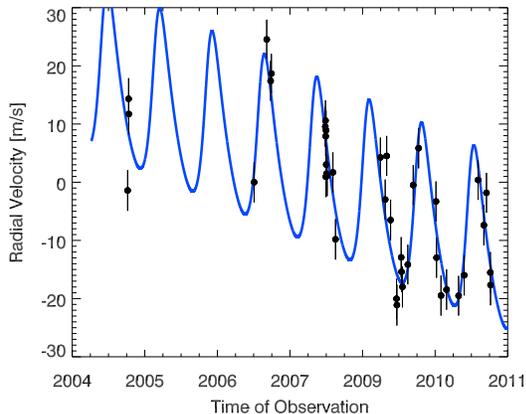,width=1.1\linewidth,clip=}
\caption{The precise Doppler velocities for HD~164509 are shown here as the black circular points with associated 1 $\sigma$ uncertainties. The solid curve is the theoretical Keplerian model that best fits this system. }
\label{figHD164509rv}
\end{figure}

\begin{figure}
\epsfig{file=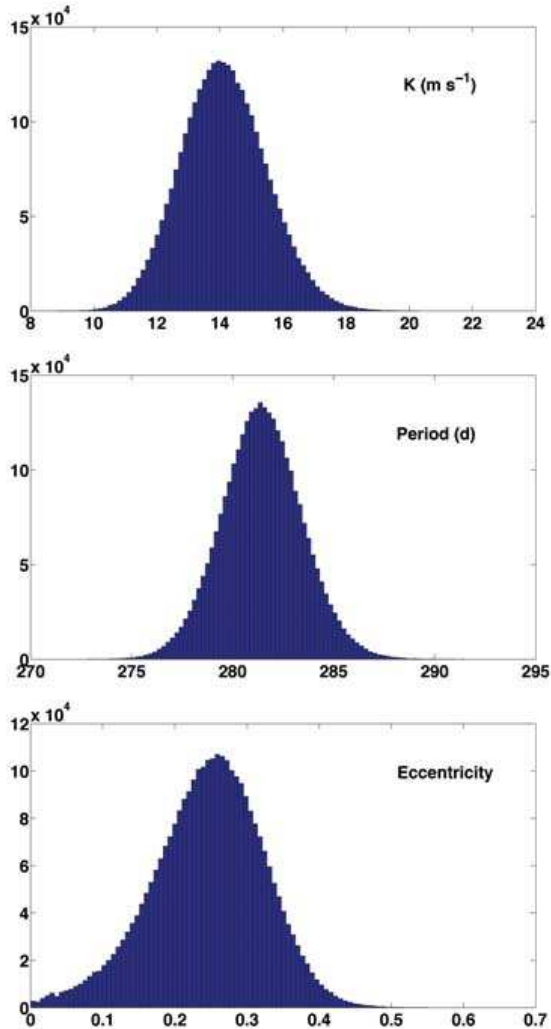,width=1.1\linewidth,clip=}
\caption{MCMC posterior PDFs for several orbital parameters for HD~164509b.}
\label{fig:bayes164}
\end{figure}

Our best-fit model is for a single planet with a period of 282.4 \plm 3.8 days, orbital eccentricity of 0.26 \plm 0.14, a radial velocity amplitude of 14.2 \plm 2.7 \ms\, and a linear trend of -5.1 \plm 0.7 \msy. With the assumed stellar mass of 1.13 \plm 0.02 \msun, we derive a mass for the planet of \msini = 0.48 \plm 0.09 \mjup.  The rms to our model fit is 4.9 \ms; with the assumed jitter, the \chinu\ for this fit is 2.04, suggesting that either the jitter has been underestimated or there are additional weak dynamical signals that have not been adequately modeled with a single planet fit. Figure \ref{figHD164509rv} shows the best fit Keplerian model for the data and the orbital parameters are summarized in Table \ref{tab:orb}. Similar to the HD~163607 system, we carried out MCMC simulations following \citet{Hou2011}. The resulting posterior PDFs can be seen in Figure \ref{fig:bayes164}, which were again in excellent agreement with the best fit solution from \textit{KFME}. Just as with HD~163607, an FAP analysis was carried out for HD~164509b. Out of the 10$^{4}$ synthetic data sets created, not a single set had a lower \chinu fit than the unscrambled velocities, leading to an FAP of $\le 0.01 \%$. The right-most plot in Figure \ref{fig:FAPs} shows the \chinu distribution for the 10$^{4}$ synthetic data sets created to estimate the FAP of HD~164509b, and the downward pointing arrow shows the \chinu for the best fit to the unscrambled velocities. 

\subsection{Photometric Observations}
\label{sec:pho164}
Results for HD~164509 are given in the third row of Table \ref{tab:pho}.  These observations were acquired between 2007 October 26 and 2010 October 9. The standard deviation of the comparison star observations is 0.0018 mag, consistent for constant stars.  Periodogram analysis of these $C2-C1$ comparison star observations finds no significant periodicity.  We compute the $P_{g}-C1C2$ differential magnitudes as we did for HD~163607 above.  The resulting low standard deviation of 0.0018 mag, along with the absence of any significant periodicity in the $P_{g}-C1C2$ observations, demonstrates that HD~164509 is constant to high precision. 

A least-squares sine fit of the $P_{g}-C1C2$ observations phased to the 282.4-day RV period gives a semi-amplitude of $0.0006\pm0.0002$ mag.  
This tight limit to brightness variability further supports the existence of HD~164509b.  Again, our observations are too few for any transit search. 

\section{Summary \& Discussion}
\label{sec:disc}

In this paper, we present Doppler velocities for two metal rich stars from the N2K survey:  HD~163607 and HD~164509. The velocities for the G5 IV star, HD~163607 are well-fit with a two-planet Keplerian model with \chinu =  1.03 and residual rms of 2.9 \ms. Based on the 75 and 1314 day orbital periods and semi-amplitudes listed in Table \ref{tab:orb}, the derived minimum masses are M$_{P}\ \sin{i}$ = 0.77 \mjup\ and M$_{P}\ \sin{i}$ = 2.29 \mjup, respectively.  \citet{if10} estimate the jitter for a subgiant star of this color to be 4.3 \ms, which gives  a \chinu ~of 0.4. However, Hipparcos data show that this star is only modestly evolved off of the main sequence. Using the \citet{if10} jitter for a main sequence star of this color of 2.6 \ms ~gives a \chinu ~of $\sim$ 1, indicating this lower value is a much better estimate of the uncertainty, which is why we chose to use that lower estimate for the uncertainty for this work. 

The velocities for HD~164509 are best fit by a single Keplerian model with a period of 282 days and a linear trend. The planet is in an orbit with modest eccentricity, $e = 0.26$, and has a minimum mass, \msini = 0.48 \mjup. However, the Keplerian model has a \rms\ of 4.9 \ms\ and \chinu\ of 2.04, suggesting that the model does not fully describe our data.  In this case, the stellar jitter may have underestimated astrophysical noise, or an additional companion is contributing to the residual velocities. 

The HD~163607 system is particularly interesting for a number of reasons. The inner planet, with an eccentricity of 0.73, is the most eccentric planet in a multiplanet system.  The average eccentricity of planets in multiplanet systems is 0.22 \citep{Wright2009}. It is not clear why this system harbors such a high eccentricity inner planet. Could the inner planet have been scattered inwards by the outer planet? Or possibly a third planet that was ejected from the system? Is the Kozai mechanism responsible, due to the outer planet or a distant stellar companion \citep{Nagasawa2008}? Are we seeing a high-eccentricity snapshot of a system that oscillates by dynamical interactions between low and high eccentricity states?

Because of the high eccentricity (e = 0.73) and orbital configuration with respect to the Earth ($\omega$ = 79$^{\circ}$) of HD~163607b, the probability of transit for the inner planet is much higher than it would have been if the orbit was circular. Planet-planet scattering simulations suggest that highly eccentric systems might also have large spin-orbit misalignments, yet there are only a handful of known systems to test this theoretical result \citep{2008ApJ...686..580C,2008ApJ...686..603J} .  If HD~163607b were to transit, spectroscopic observations taken during transit for this bright and highly eccentric system would allow for the calculation of the spin-orbit alignment \citep{qhs+2001, 2005ApJ...622.1118O, 2009ApJ...703.2091W}, making this an excellent system to test theoretical predictions. While we have attempted to detect a transit event for the inner planet orbiting HD~163607, the combination of the orbital period being relatively long and the orbital period being a non-integer multiple of a day have made these attempts quite difficult. We encourage members of the community to search for transit events of this inner planet.

\acknowledgements
We gratefully acknowledge the dedication and support of the Keck Observatory staff, in particular Grant Hill and Scott Dahm for their support with HIRES and Greg Wirth and Bob Kibrick for supporting remote observing. We thank the NASA Telescope assignment committees for generous allocations of telescope time. We thank the anonymous referee for useful comments and suggestions. Fischer acknowledges support from NASA grant NNX08AF42G and NASA Keck PI data analysis funds. GWH acknowledges support from NASA, NSF, Tennessee State University, and the State of Tennessee through its Centers of Excellence program.  JTW gratefully acknowledges the Center for Exoplanets and Habitable Worlds, supported by the Pennsylvania State University, the Eberly College of Science, and the Pennsylvania Space Grant Consortium. Data presented herein were obtained at the W. M. Keck Observatory from telescope time allocated to the National Aeronautics and Space Administration through the agency's scientific partnership with the California Institute of Technology and the University of California.  The Observatory was made possible by the generous financial support of the W. M. Keck Foundation.  The authors extend thanks to those of native Hawaiian ancestry on whose sacred mountain of Mauna Kea we are privileged to be guests.  Without their generous hospitality, the Keck observations presented herein would not have been possible.

%%%%%%%%%%%%%%%%%%%%%%%%%%%%%%%%%%%%%%%%%%%%%%%%
%%%                                                                 TABLES                                                                         %%%
%%%%%%%%%%%%%%%%%%%%%%%%%%%%%%%%%%%%%%%%%%%%%%%%

%%%%%%%%%%%%%%%%%%%%%%%%%%%%%%%%%%%%%%%%%%%%%%%%
%%%                                                                 FIGURES                                                                       %%%
%%%%%%%%%%%%%%%%%%%%%%%%%%%%%%%%%%%%%%%%%%%%%%%%

\clearpage

\clearpage
\end{document}